\documentclass[12pt]{article}
\usepackage{graphics,cite,amssymb,epsfig,float,psfrag}
\usepackage[usenames,dvips]{color}
\usepackage{rotating}

\begin{document}

\newcommand{\lsim}{\raisebox{-0.13cm}{~\shortstack{$<$ \\[-0.07cm] $\sim$}}~}
\newcommand{\gsim}{\raisebox{-0.13cm}{~\shortstack{$>$ \\[-0.07cm] $\sim$}}~}
\newcommand{\ra}{\rightarrow}
\newcommand{\lra}{\longrightarrow}
\newcommand{\ee}{e^+e^-}
\newcommand{\gam}{\gamma \gamma}
\newcommand{\nn}{\noindent}
\newcommand{\non}{\nonumber}
\newcommand{\beq}{\begin{eqnarray}}
\newcommand{\eeq}{\end{eqnarray}}
\newcommand{\s}{\smallskip}
\baselineskip=14.8pt
%%%%%%%%%%%%%%%%%%%%%%%%%%%%%%%%%%%%%%%%%%%%%%%%%%%%%%%%%%%%%%%%%%%%%%%%

\vspace*{4mm}
\rightline{LPT--Orsay 12/31}
\rightline{CERN-PH-TH/2012--087}
%\rightline{hep-ph/0610173}
%\rightline{October 2006}   

\vspace{0.5cm}

\begin{center}

{\large {\bf Sealing the fate of a fourth generation of fermions}}

\vspace*{1cm}

{\sc Abdelhak Djouadi$^{1,2}$ and Alexander Lenz$^{2}$} 

\vspace{0.97cm}

$^1$ Laboratoire de Physique Th\'eorique, CNRS and Universit\'e Paris--Sud, \\
B\^at. 210, F--91405 Orsay Cedex, France.

\vspace{0.2cm}

$^2$ Theory Unit, Department of Physics, CERN, CH-1211 Geneva 23, Switzerland.

\end{center}

\vspace{0.5cm}

\begin{abstract} 
\noindent  The search for the effects of heavy fermions in the extension of the
Standard Model with a fourth generation is part of the experimental program of
the  Tevatron and LHC experiments. Besides being directly produced, these states
affect  drastically the production and decay properties of the Higgs boson.  In
this note, we first reemphasize the known fact that in the case of a light  and
long--lived fourth neutrino, the present collider searches do not permit to
exclude a Higgs boson with a mass below the $WW$ threshold.  In a second step, 
we show that  the recent results from the ATLAS and CMS collaborations  which
observe an excess in the $\gamma \gamma$ and $4\ell^\pm$ search channels
corresponding to a Higgs boson with a mass $M_H\! \approx\! 125$ GeV, cannot
rule  out the fourth generation possibility if the $H\!\to \!\gamma \gamma$
decay rate is evaluated  when naively implementing the leading  ${\cal O}(G_F
m_{f'}^2)$ electroweak corrections. Including the exact next-to-leading order
electroweak corrections  leads to a strong suppression of the  $H\!\to\! \gamma
\gamma$ rate and makes this channel unobservable with present data. Finally, we
point out that the observation by the Tevatron collaborations of a $\gsim
2\sigma$ excess in the mass  range $M_H\!=\! 115$--135 GeV in the channel $q\bar
q\! \to\! WH\! \to\! Wb\bar b$ can definitely not be accommodated by the fourth
gene\-ration fermion  scenario.  All in all, if the excesses observed at the LHC
and the Tevatron are indeed due to a Higgs boson, they unambiguously exclude the
perturbative fermionic fourth  generation case. In passing, we also point out
that the Tevatron excess definitely rules out the fermiophobic Higgs scenario as
well as  scenarios  in which the Higgs couplings to gauge bosons and bottom
quarks are significantly reduced.  
\end{abstract}

\newpage

One of the most straightforward  extensions of the Standard Model (SM) of
particle physics is to assume a fourth generation of fermions: one simply adds
to  the known fermionic pattern with three generations, two quarks $t'$ and $b'$
with weak--isospin of respectively $\frac 12$ and  $-\frac 12$, a charged lepton
$\ell'$ and a neutrino $\nu'$. Such an extension that we will denote by SM4,  
besides of being rather simple, has been advocated as a possible solution of
some problems of the SM; for recent reviews and motivations for a fourth fermion
generation, see Refs.~ \cite{Frampton,Kribs,Holdom,Flavor}. For instance, from a
theoretical point of view,  it provides new sources of CP--violation that could
explain the  baryon asymmetry in the universe \cite{Holdom} and, from the
experimental side, it might soften some tensions in flavour physics
\cite{Flavor}.\s 

There are, however, severe constraints on this SM4 scenario. First, from  the
invisible width of the $Z$ boson, the LEP experiment has measured the number of
light neutrinos to be  $N_\nu=3$ with a high precision \cite{PDG} and, thus, the
neutrino of SM4 should be rather heavy, $m_{\nu'} \gsim \frac12 M_Z$,  assuming
that it has a very small mixing with the lighter SM leptons (not to be produced 
in association with its light partners which would lead to the stronger  limit 
$m_{\nu'} \gsim 100$ GeV). A heavy charged lepton with a mass   $m_{\ell'}\!
\lsim\! 100$ GeV has also been excluded at LEP2 \cite{PDG}.  In addition, the
Tevatron and now the LHC experiments have excluded too light fourth generation
quarks. In particular, direct searches performed by the ATLAS and CMS
collaborations rule out  heavy down-type and up--type quarks with  masses
$m_{b'} \lsim 600$ GeV and $m_{t'} \lsim 560$ GeV  \cite{LHC4}. On the other
hand, high precision electroweak data severely constrain the mass splitting
between the fourth generation quarks while data from B--meson physics constrain
their mixing pattern \cite{Constraints}.  Finally, the requirement that SM4
remains unitary at very high energies suggests that fourth generation fermions
should not be extremely heavy, $m_{q'} \lsim 500$ GeV  \cite{Unitarity}. 
However, this bound should not be viewed as a strict limit but simply as an
indication that strong dynamics takes place; a degenerate quark doublet with a 
mass $m_{q'} \approx 700$ GeV has been considered in a simulation of  a strong
Yukawa coupling regime on the lattice  \cite{Lattice}.  Thus,  ATLAS and CMS
direct searches  for $t',b'$ SM4 quarks are closely approaching the masses 
required by  the perturbative unitarity bound and   we will assume here that
$m_{t'}\! \approx\!  m_{b'}\! \pm \!50\;{\rm GeV}\! \sim\! 650\;$GeV.\s

Strong constraints on SM4 can be also obtained from Higgs  searches at the
Tevatron and the LHC. Indeed, it is known since a long time \cite{history} that
in the loop induced Higgs--gluon and Higgs-photon vertices, $Hgg$ and
$H\gamma\gamma$, any heavy particle coupling to the Higgs boson proportionally
to its mass, as is the case in SM4, will not decouple from  the amplitudes and
would have a drastic impact.  In particular, for the $gg\to H$ process
\cite{ggH-LO,reviews} which is the leading mechanism  for  Higgs production at
both the Tevatron and the LHC, the additional contribution of the  two new SM4 
quarks  $t'$ and $b'$ will increase the rate by  a factor of $K_{gg\!\to\!
H}^{\rm SM4} \approx 9$. At leading order in the electroweak interaction, this
factor is a very good approximation \cite{Hgg-SM4}, as long as the heavy quarks
are such that $m_{q'} \gsim \frac12 M_H$ which holds true for any $M_H$ value
below the TeV scale, given the experimental bounds on the $q'$ masses. The Higgs
searches at the Tevatron and the LHC, which are now becoming very sensitive,
should therefore   severely constrain the SM4 possibility \cite{earlier}.
Indeed, the CDF and D0 experiments for instance exclude a Higgs boson in this
scenario for masses $124\;{\rm GeV}\! \lsim \! M_H \lsim 286\;$GeV by
considering mainly the $gg\! \to\! H \!\to  \! WW \! \to\!  2\ell 2\nu$ channel
\cite{SM4-TeV}. The LHC experiments recently extended this exclusion  limit up
to $M_H\! \approx \!600$ GeV (at 99\%~CL)  by exploiting also the $gg \! \to  \! H \!
\to ZZ \! \to \! 4\ell, 2\ell 2\nu, 2\ell 2j$ search channels \cite{SM4-LHC}.\s

Nevertheless there are  two caveats which might loosen these experimental
limits. The first one is that the electroweak radiative corrections to the
$gg\to H$ process turn  out to be significant
\cite{ggH-NLOapprox,EW-exact,4thpaper}. For a specific choice of fermion masses
which approximately fulfills the electroweak precision constraints 
\cite{Kribs}, $m_{b'}\! = \! m_{t'}+50\;{\rm GeV}= \! m_{\ell'} \! = \! m_{\nu'}
\sim 600 $ GeV,  they lead to an increase (decrease) of the cross section at low
(high) Higgs masses, $M_H\!\approx\! 120\;(600)$ GeV, by $\approx 12\%$ implying
that the  exclusion limits above need to be updated and changes in the excluded
$M_H$  range up to 10 GeV are expected \cite{4thpaper}.\s

The second  caveat   is that the Tevatron and LHC Higgs exclusion limits in SM4 
are only valid for a heavy neutrino $\nu'$. Indeed, if $m_{\nu'} \lsim \frac12
M_H$, the Higgs boson will also decay into a neutrino pair \cite{RV} and the
branching ratio BR($H\to \nu' \bar \nu'$) can be sizable enough to suppress the
rates for the visible channels such as $H\to WW,ZZ$ by which the Higgs is
searched for. This is particularly the case for a light Higgs, $M_H \lsim 160$
GeV, which mainly decays into $b$--quark pairs and $W$ bosons (with one   $W$
being virtual). The Higgs total width is small in this case, $\Gamma_H \lsim 1$
GeV,  making the invisible channel  $H\!\to\! \nu' \bar \nu'$   dominant.\s 

Using the program {\tt HDECAY} \cite{HDECAY} in which the Higgs decays in SM4,
with all  known  QCD \cite{Hgg-SM4} as well as the leading ${\cal O}(G_F
m_{f'}^2)$ electroweak  and ${\cal O}(G_F  m_{q'}^2\alpha_s)$ mixed  
corrections  derived  in Ref.~\cite{EW-approx} have been (naively)
implemented\footnote{For the $H \to gg, f\bar f$ and $VV$ decays, the ${\cal
O}(G_F m_{f'}^2)$ terms when implemented by simply multiplying the couplings
$g_{HXX}$ by the electroweak correction $1+\delta_{EW}^X$, should represent a
good approximation  \cite{EW-proph}. A fourth generation of fermions   with
degenerate $t',b',\ell', \nu'$ masses $m_{f'} \approx  300\;(600)\;$GeV will
suppress the $HVV$ coupling by $\approx 10\%\;(40\%)$ and, hence, the rate for
the  $H\to VV$ decay (which grows like the square of the coupling) by $20\%\; 
(80\%)$ \cite{EW-approx}.  However, in the $H\gamma\gamma$ amplitude, this
approximation  leads to an unstable result and some reordering of the 
perturbative series is needed \cite{4thpaper}  as will be  discussed later.}, we
exemplify this feature  in  Fig.~1 where the Higgs decay branching ratios into
$VV$ states normalised to their SM values, BR$(H\! \to\! VV)|_{\rm SM4/SM}$, 
are shown as a function of $m_{\nu'}$ for  $M_H\!=\!125\;$GeV (with this
normalisation, these ratios are the same for $V\!=\!W$ and $Z$). One first
observes that for $m_{\nu'} \gsim \frac12 M_H$, the ${\cal O}(G_F m_{\ell'}^2)$
corrections suppress the rate  for $H \!\to\! VV$ decays  while they increase
the one for the $H\to \gamma \gamma$ channel.  In addition, one can see that for
a heavy neutrino $\nu'$,  say $m_{\nu'}=300$ GeV, BR$(H\to WW,ZZ)$ are  
suppressed by only a factor of $\approx 5$ compared to their SM values, as a
result of the additional $t',b'$ contributions. However,  when the $H\to
\nu'\bar \nu'$ decay channel is kinematically allowed, i.e. for $\frac12 M_Z
\lsim m_{\nu'} \lsim \frac12 M_H$, BR$_{\rm SM4/SM}$ is further suppressed and
for a given neutrino mass, the suppression factor is comparable to or even
larger than the factor $\approx 9$ due to the increase of the $gg\to H$ cross
section by the $t',b'$ loop contributions. Thus, the rate for the processes
$gg\!\to \! H\! \to \! WW,ZZ$ can be smaller in SM4 compared to the SM and,
hence, the Tevatron and LHC exclusion limits  of  Refs.\cite{SM4-TeV,SM4-LHC},
which are obtained using these processes, can be evaded\footnote{Note that  for
larger Higgs mass values, $M_H \gsim 180$ GeV, the  $H\to WW,ZZ$ partial widths
become large and these decays are by far dominant and  thus not affected by the
presence of the $H\to \nu'\bar \nu'$ channel. The present Tevatron and LHC
exclusion limits are valid in this case, modulo the impact of the electroweak 
corrections to the production and decay processes which need to be included.}.\s

\begin{figure}[!h]
\vspace*{1mm}
\hspace*{-2.5cm} 
\epsfig{file=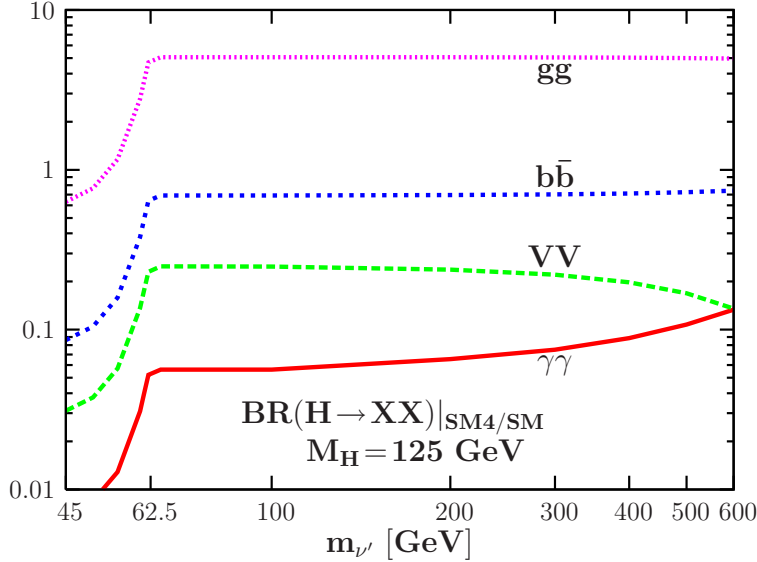,width=11.5cm} 
\vspace*{-14.7cm} 
\caption[]{\small The decay branching ratios of a 125 GeV Higgs particle
into $gg,b\bar b, \gamma\gamma$ and $VV$ states (with $V\!=\!W,Z$) in 
SM4  normalised to their SM values as a function of the  neutrino 
mass. The heavy quark masses are set to $m_{b'}\!=\!m_{t'}\!+\!50~{\rm GeV}
\!=\!600$ GeV, while the charged lepton mass is  $m_{\ell'}\!=\!m_{\nu'}\!+
\!50~{\rm GeV}$. The  electroweak corrections are  included in a naive way 
in $H\to \gamma\gamma$.}
\label{BRSM4}
\vspace*{-3mm}
\end{figure}

Let us now discuss, in the context of SM4, the excess of events recently
observed by the ATLAS and CMS collaborations  in the $H \! \to \! ZZ \! \to \!
4\ell^\pm$ and $H\!\to \! \gamma \gamma$ channels corresponding to a SM--like
Higgs boson with $M_H \! \approx \!125$ GeV \cite{evidence} . First of all,
for the value $M_H= 125$ GeV, while BR$(H\! \to\! ZZ)|_{\rm SM4/SM}$ is
different from unity as a result of the ${\cal O}(G_F m_{f'}^2)$ corrections,
the  enhancement of the $H\to gg$ rate by the $t',b'$ contributions  and
eventually the opening of the $H\!\to \! \nu' \bar \nu'$ mode, the situation is
more complicated in the case of BR$(H\! \to\! \gamma\gamma)|_{\rm SM4/SM}$ as
there is another  important effect. As a matter of fact,  the
$H\!\to\!\gamma\gamma$ decay is mediated by $W$ boson and heavy fermion loops
whose contributions interfere destructively. While this interference is mild in
the SM, as the $W$ contribution is much larger than that of the top quark, it is
very strong in  SM4 because of the additional $t',b'$ and $\ell'$
contributions; the $W$ and all fermion contributions are then very close to each
other but opposite in sign.  This accidental cancellation makes  BR$(H\!\to\!
\gamma \gamma)|_{\rm SM4/SM}$  much smaller than BR$(H\! \to\! VV)|_{\rm
SM4/SM}$ in general, with consequences  summarized below.\s 

It is clear from Fig.~1 that in the presence of a relatively light neutrino, 
$m_{\nu'} \lsim \frac12 M_H$, the rates for the $H\to ZZ$ and $H\to \gamma
\gamma$ decays are  strongly suppressed by a factor that is larger than the one
$K_{gg\!\to\! H}^{\rm SM4}\! \approx\! 9$ which enhances the $gg\!\to\! H$ cross
section. Thus, the $\gamma \gamma$ and $4\ell^\pm$ excesses corresponding  to a
SM--like 125 GeV Higgs cannot occur in SM4 when the channel $H\!\to\! \nu'
\nu'$ is open. The possibility $m_{\nu'}\! \lsim\! \frac12 M_H$ is thus  
strongly disfavored.\s 

On the other hand, when this new decay channel is closed, the  rates for $H\!\to
\!VV,\gamma\gamma$ decays  increase significantly.  If one assumes heavy leptons
with $m_{\nu'}\!\approx \! m_{\ell'}\!\approx \!600$ GeV (and taking into
account the fact that the electroweak corrections decrease BR$(H\!\to\! VV)$
with increasing $m_{\nu'}$), one  accidentally obtains a suppression rate
BR$_{\rm SM4/SM} \approx 7.5$ that is the same in both cases. Recalling  that 
in this case the rate for the main Higgs production process is enhanced by a
factor  $K_{gg\!\to\! H}^{\rm SM4}\! \approx\! 9.5$, one obtains  $gg\! \to \!H
\! \to\! \gamma\gamma$ and $4\ell$ rates for $M_H=125$ GeV that are a $\approx
20\%$ larger in SM4 than in the SM (see also Ref.\cite{Hgamma}). It happens 
that the excesses observed by ATLAS and CMS in the $\gamma \gamma$ channel  are
stronger than what is expected in the SM,  although within the errors bands.
Therefore, not only a fourth generation with all heavy fermions having a mass
close to $m_{f'}\!=\!600$ GeV could accommodate the excesses observed at the
LHC, but it could also explain the  substantial  rate observed by ATLAS and CMS
in the $H\to \gamma\gamma$ signal that has the largest significance.\s

There is, however, a serious flaw in the discussion above. As mentioned 
earlier, only the leading ${\cal O}(G_Fm_{f'}^2)$ terms  (and the  ${\cal
O}(G_Fm_{q'}^2 \alpha_s)$ ones)   \cite{EW-approx} are included in the 
electroweak corrections  to  BR($H\to \gamma\gamma$) in  Fig.~1, by simply
multiplying the $W,t$ and $f'$ amplitudes with the relevant correction
$1+\delta_{EW}^X$. The exact next-to-leading order (NLO) electroweak 
corrections have been very recently calculated \cite{4thpaper} and, because  of
the very strong interference  between the $W$ and all fermion loop
contributions,  they have a drastic impact on the $H\gamma \gamma$ vertex.  For
$m_{f'}\! \approx\! 600$ GeV, these corrections suppress BR($H\!\to\!\gamma
\gamma) |_{\rm SM4}$ by  almost an order of magnitude, compared to the case
where the ${\cal O}(G_Fm_{f'}^2)$ corrections  are naively implemented in the
amplitudes. Nevertheless, it has been shown  \cite{4thpaper} that by reordering
the  perturbative series and including subleading $M_H^2/4M_W^2$ terms in the
$W$ amplitude, one can reproduce the relative NLO electroweak corrections of the
exact result at the percent level. An adapted version of the program {\tt
HDECAY} implements  this  approximation of the full NLO electroweak corrections
to the decay $H\to \gamma\gamma$  in SM4  \cite{new-hdecay}.\s 

Using this new version of  {\tt HDECAY}, we display in the left--hand side of 
Fig.~2  the cross section times branching ratio  $\sigma(gg\! \to\! H)\!\times\!
{\rm BR}(  H\! \to\! \gamma \gamma)|_{\rm SM4/SM}$ at $M_H\!=\!125$ GeV as a
function of $m_{\nu'}\!=\!m_{\ell'}$ for the value
$m_{b'}\!=\!m_{t'}\!+\!50\!=\!600\;$GeV (the change when varying $m_{b'}$ in the
still allowed range 600--700 GeV should be mild). As can be seen,  $\sigma(gg\!
\to\! H)\!\times\! {\rm BR}(  H\! \to\! \gamma \gamma)$ in SM4 is a factor of 5
to 10 smaller than in the SM. The increase of $\sigma(gg\!\to\! H)$ by a factor
of $\approx 9.5$  in SM4 is thus not  sufficient  for the $\gamma \gamma$ signal
to be observed by the ATLAS and CMS experiments, hence  excluding the
perturbative SM4 scenario if the $\gamma \gamma$ and $4\ell^\pm$ excesses  at
the LHC are indeed due to a 125 GeV Higgs boson.\s

\begin{figure}[!h]
\vspace*{-8mm}
\vspace*{-2cm}
\mbox{\hspace*{-6cm} \epsfig{file=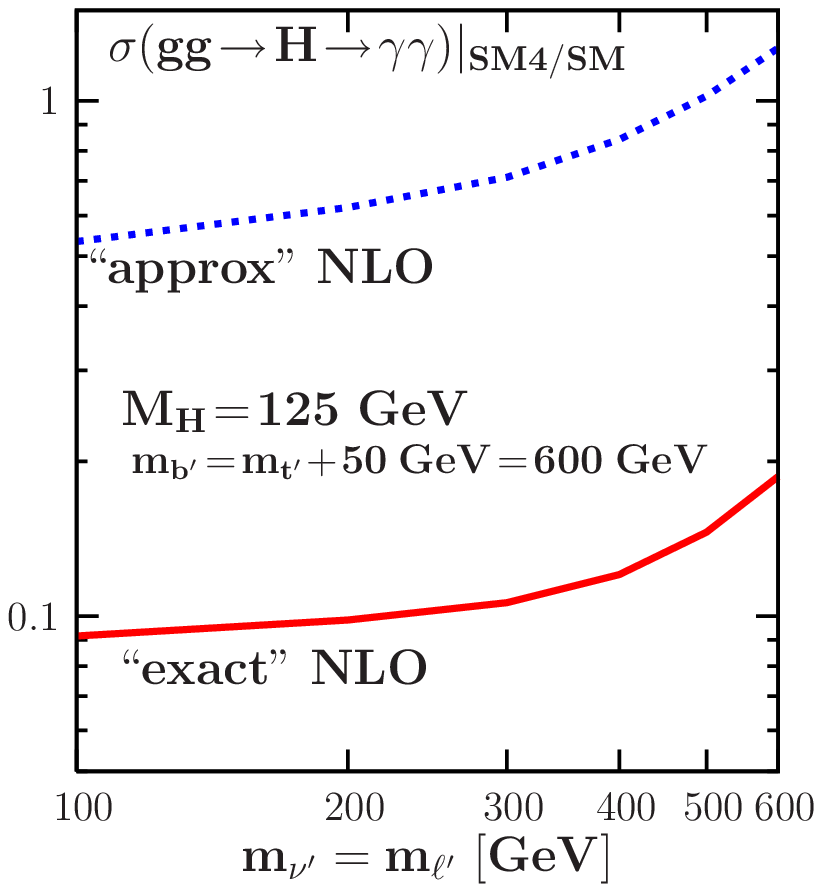,width=18cm}
       \hspace*{-10cm}\epsfig{file=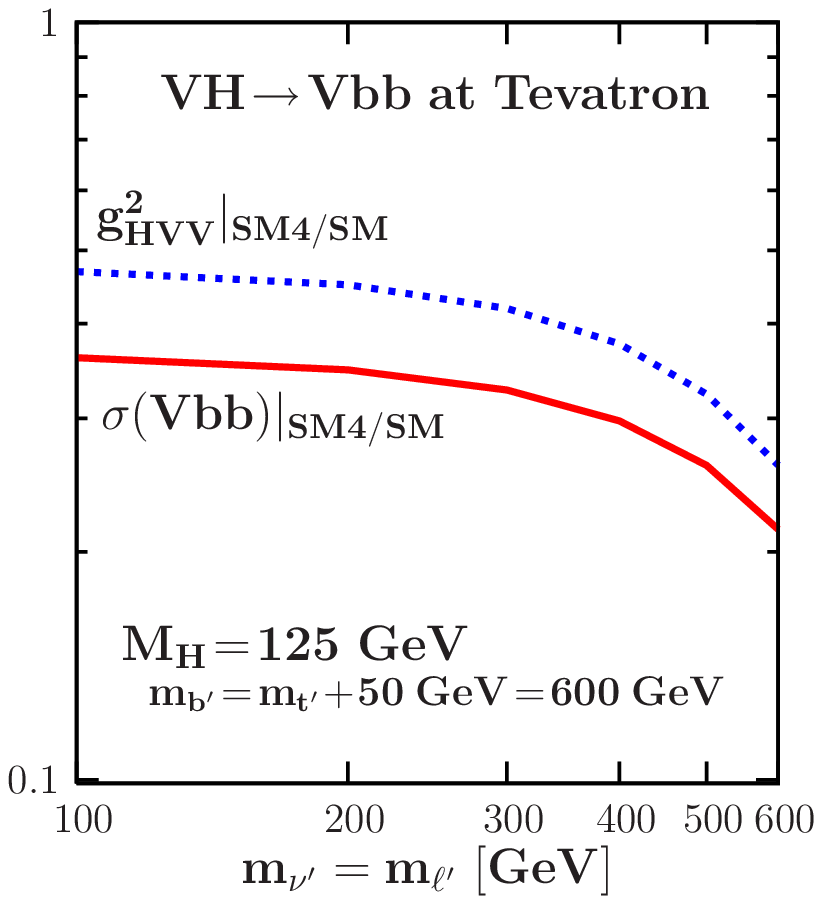,width=18cm}
} 
\vspace*{-15.6cm} 
\caption[]{\small Left: $\sigma(gg\! \to\! H)\!\times\! {\rm BR}(  H\! \to\!
\gamma \gamma)|_{\rm SM4/SM}$ for a $125$ GeV Higgs boson as a function of
$m_{\nu'}\!=\!m_{\ell'}$ when the leading ${\cal O}(G_Fm_{f'}^2)$ electroweak 
corrections are  included in a naive way  (``approx" NLO) or in a way that mimics
the exact  NLO results (``exact" NLO). Right:  the $HVV$ coupling squared and
$\sigma(q\bar q\! \to\! VH)\!\times\! {\rm BR}(H\!\to\! b\bar b)$ in SM4 
normalized to the SM values.} 
\label{BRSM4-gg}
\vspace*{-1mm}
\end{figure}

A final argument against the existence of a fourth generation, and which is 
theoretically more robust than the argument above based on the LHC $\gamma
\gamma$ signal that is subject to large cancellations in the $H\to \gamma
\gamma$ amplitude, is provided by the recently updated SM Higgs search  by the
CDF and D0  collaborations  with up to 10 fb$^{-1}$ of data \cite{Tevatron}. In
this  search, a $\approx 2.2 \sigma$ excess of data has been observed in the
Higgs mass range between  115 and 135 GeV and is mostly concentrated in the
Higgs--strahlung channel $q\bar q\! \to\! VH \to\! V b\bar b$ with $V\!=\!W,Z$;
this excess thus strengthens the  case for a $\approx\! 125$ GeV Higgs  boson at
the LHC.

\noindent In SM4, such an excess  cannot occur for the following two reasons.
First, compared to the SM, the $HVV$ coupling and hence  the production cross
sections $\sigma(q\bar q \to VH) \propto g_{HVV}^2$ are strongly suppressed by 
the leading ${\cal O}(G_Fm_{f'}^2)$ corrections (which  approximate well  the
full electroweak NLO corrections in this case \cite{EW-proph}) as mentioned
earlier. Second, the branching ratio BR$(H\! \to \! b\bar b)$ in SM4 is
significantly affected by the presence of the new $t',b'$ quarks and,  as shown
in Fig.~1, is $\approx 30\%$ smaller than  in the SM for $M_H\! \approx\!
125$ GeV. \s

The ratio $\sigma(q\bar q\! \to\! VH)\!\times\! {\rm BR}(  H\! \to\! b\bar
b)|_{\rm SM4/SM}$ is thus much smaller than unity as exemplified in Fig.~2
(right) where it is displayed as a function of $m_{\nu'}\!=\!m_{\ell'}$ again
for $m_{b'}\!=\!m_{t'}\!+\!50\!=\!600$ GeV. This reduction of the $Vb\bar b$
signal rate by a factor 3 to 5 depending on the  $m_{\nu'}$ value would make the
Higgs signal unobservable at the Tevatron and, therefore,  the $2.2$ excess seen
by CDF and D0, if  indeed due to a $\approx 125$ GeV Higgs boson,  unambiguously
rules out  the SM4 scenario with perturbative  Yukawa couplings.\smallskip

\begin{figure}[!h]
\vspace*{-2.6cm}
%\hspace*{-2.5cm} 
\epsfig{file=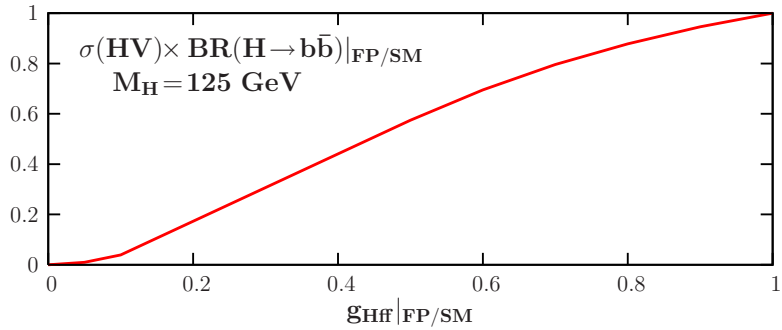,width=15cm} 
\vspace*{-14.7cm} 
\caption[]{\small The rate $\sigma(q\bar q \! \to \! VH) \! \times \! {\rm BR}
(H\! \to \!  b\bar b)$ of a 125 GeV Higgs boson normalised to its SM  value as 
a function of the ratio  $g_{Hff}|_{\rm FP/SM}$ of couplings in a  fermiophobic
Higgs scenario. The program {\tt HDECAY} \cite{HDECAY}, in which the 
fermiophobic Higgs scenario is implemented, has been used.}
\label{FP}
\vspace*{-1mm}
\end{figure}

Finally, one should note that the observation of the channel $q\bar q\to VH$
with $H \to b \bar b$ at the Tevatron  would also definitely exclude the
fermiophobic Higgs scenario. This possibility has been advocated to explain the
excess of events  at the LHC in the channel $\gamma \gamma$  (plus additional
jets), although the  fit probability is not larger than in the SM 
\cite{fermiophobic}. The observation of the Tevatron excess in $\ell \nu b\bar
b$ events can occur only if the decay $H\to b\bar b$ is present. In fact, even
cases in which the $Hb\bar b$ coupling is non--zero but suppressed compared to
its SM value are disfavored. Indeed, as the rate $\sigma (WH)\! \times \!BR(H
\!\to \! b \bar b)$ is, to a good approximation, $\propto g_{HWW}^2 \! \times \!
\Gamma(H \! \to \! b\bar b) / [\Gamma(H\! \to \! b\bar b)\! + \! \Gamma(H \! 
\to \! WW^*)]$ with $\Gamma(H\! \to \! WW^*, b\bar b) \propto g_{HWW, Hbb}^2$, a
suppression by 10\%, 50\% and 90\% of the couplings $g_{Hff}$ would lead to a
suppression of $\sigma(Wb\bar b)$  by, respectively, $\approx 5\%$, 42\% and
96\%. This is exemplified in Fig.~\ref{FP} in which the cross section times
branching ratio $\sigma(q \bar q \to VH) \times {\rm BR}(H\to b\bar b)$,
normalized to its SM value, is displayed when the fermionic Yukawa couplings
$g_{Hff}|_{\rm FP/SM}$ are collectively varied  from  zero (i.e. the pure
fermiophobic case) to unity (the SM case).\s

This argument can be extended to many models in which either the $HVV$ coupling 
or the $H\to b\bar b$ branching fraction (or both) are significantly suppressed
compared to their SM values as could be the case in, for instance, minimal
composite Higgs models \cite{composite}.\s

In summary, we have pointed out that while the exclusion bounds on a light Higgs
boson, $M_H\lsim 160$ GeV, in  SM4 with a fourth generation can be evaded by 
assuming a light fourth neutrino, $m_{\nu'} \lsim \frac12 M_H$, this
possibility  is excluded by the  observation at the LHC of a Higgs signal at a
mass $\approx 125$ GeV  in the $\gamma \gamma$ and $ZZ^*\to 4\ell^\pm$ final
states.  The  ATLAS+CMS $4\ell^\pm$ and $\gamma \gamma$ signals  are compatible
with the SM4 scenario (and larger signal rates than in the SM could even be
accommodated)  if  the leading ${\cal O}(G_F m_{f'}^2)$ electroweak corrections
are naively included in the $H \to \gamma\gamma$ rate. However, when  including 
the full set of electroweak  corrections at next-to-leading order
\cite{4thpaper}, the $\gamma \gamma$  signal is suppressed  by an order of
magnitude compared to the previous approximation, hence strongly disfavoring a
perturbative SM4.   Finally, the observation by the CDF/D0 collaborations of a
$\gsim\! 2\sigma$ excess corresponding to a Higgs boson with $M_H\!=\! 115$--135
GeV in the channel $q\bar q\! \to\! VH\! \to\! Vb\bar b$ can definitely not be
accommodated in  SM4.  Hence, if the excesses observed at the LHC and the
Tevatron are indeed the manifestations  of a $125$ GeV  Higgs boson, the
scenario  with a perturbative fourth fermionic generation is unambiguously
excluded.\s

{\it En passant}, we also point out that the pure fermiophobic Higgs scenario
cannot accommodate the Higgs signal in the $VH \to Vb\bar b$ channel observed at
the Tevatron. In fact, many scenarios in which the $Hb\bar b$ or $HWW$ couplings
(or both)  are  suppressed compared to their SM values are disfavored if the 
Tevatron excess is indeed due to a Higgs particle.\bigskip

{\bf Note added:} On July 4th, 2012 ATLAS and CMS announced new results for the Higgs boson
search \cite{LHC_July}, which has severe implications on the fate of a fourth generation of fermions.
Before this announcement a combined fit of electro-weak precision data and Higgs production
and decay data yielded the result, that the SM4 is excluded by 3.1 standard deviations 
\cite{Eberhardt:2012ck}. The new data worsens the situation for the SM4 in several points: 
\begin{enumerate}
\item
Both ATLAS and CMS see a $H \to \gamma \gamma$-signal with a statistical significance 
of more than 4 standard deviations. The total observed rate was higher than expected
by the SM (a factor of $1.9 \pm 0.5$ for ATLAS and a factor of $1.56 \pm 0.43$  for CMS). 
In the SM4 one would expect instead a reduction of the rate by at least a factor 
of 5 compared to the standard model, see Fig.(\ref{BRSM4-gg}). 
Thus, both ATLAS and CMS individually see a $H \to \gamma \gamma$-signal, 
which is about 4 standard deviations away from the expectation of the SM4, which rules out the SM4.
As discussed above, the theory prediction for $H \to \gamma \gamma$ in the SM4 suffers 
from severe cancellations, so one might not want to rely on this decay channel alone.
\item
On July 2nd, 2012 also CDF and D0 updated their Higgs search \cite{TeVatron:2012cn}
in the Higgs-strahlung channel, discussed above. There the statistical significance increased
from 2.6 standard deviations in \cite{Tevatron} to 2.9 standard deviations in \cite{TeVatron:2012cn}.
At $m_H = 125$ GeV Tevatron finds a signal strength of $1.97^{+0.74}_{-0.68}$, so a little above 
the SM expectation, while the SM4 predicts values below 0.4, see Fig.(\ref{BRSM4-gg}). 
Again a stronger indication against the SM4, compared to the status of Moriond 2012.
\item
In the SM4 one would expect a sizeable enhancement of the $ H \to \tau \tau $ channel, see e.g.
\cite{Eberhardt:2012ck}, which is not observed \cite{LHC_July} in the new data. A further argument against the SM4 at the 4 $\sigma$ level. 
\end{enumerate}
It is beyond the scope of this paper to make a precise statistical statement about the exclusion of the SM4. Nevertheless, we conclude that the standard model with a perturbative 4th generation and one Higgs doublet is ruled out by this new experimental developments.

\bigskip

\noindent {\bf Acknowledgements:} We thank R. Godbole, C. Grojean,   H.-J. He
and T. Volansky for discussions. Special thanks go to Michael Spira for his work
in implementing SM4 in {\tt HDECAY} and for valuable suggestions and comments.
A.D. thanks the CERN TH Unit for hospitality and support and  A.L. is supported
by DFG through a Heisenberg fellowship.

\baselineskip=14pt
\begin{small}

\end{small}

\end{document}